\documentclass[adp,fleqn]{w-art}

\usepackage{amsmath}
\usepackage{amsfonts}
\usepackage{amssymb}
\usepackage{times}
\usepackage{w-thm}
\usepackage[]{graphicx}
\usepackage{color}
\usepackage{wrapfig}
\usepackage{bm}
\usepackage{hyperref}
\hypersetup{colorlinks,linkcolor=blue,urlcolor=blue,citecolor=blue}

\chardef\bslash=`\\ 

\newcommand{\Eq}[1]{Eq.\eqref{#1}}
\newcommand{\Eqs}[1]{Eqs.\eqref{#1}}
\newcommand{\EqsAnd}[2]{Eqs.(\ref{#1}) and (\ref{#2})}
\newcommand{\Tab}[1]{Table \ref{#1}}
\newcommand{\Fig}[1]{Fig.\ref{#1}}
\newcommand{\Sect}[1]{Section \ref{#1}}
\newcommand{\Ref}[1]{Ref.\cite{#1}}
\newcommand{\BibTitle}[1]{#1}

\newcommand{\dx}[1]{{\rm d}#1}

\newcommand{\ShSq}[1]{{\rm sinh}^2 #1 }
\newcommand{\Cth}[1]{{\rm coth}\!\left( #1 \right) }
\newcommand{\Sh}[1]{{\rm sinh}\!\left( #1 \right) }
\newcommand{\Ch}[1]{{\rm cosh}\!\left( #1 \right) }
\newcommand{\Ln}[1]{\,{\rm ln}\!\left( #1 \right) }
\newcommand{\Shs}[1]{{\rm sinh}\left( #1 \right) }
\newcommand{\Chs}[1]{{\rm cosh}\left( #1 \right) }

\newcommand{\gammab}{\tilde{\gamma}}

\newcommand{\mint}[2]{{\rm min}[#1,#2]}
\newcommand{\maxt}[2]{{\rm max}[#1,#2]}

\newcommand{\tPhi}{\tau_{\varphi}}
\newcommand{\tTem}{\tau_{T}}
\newcommand{\tTho}{\tau_{\rm Th}}
\newcommand{\tDwell}{\tau_{\rm dw}}

\newcommand{\Prob}{P}
\newcommand{\ProbD}{{\cal P}}
\newcommand{\KLen}{{\cal L}}

\newcommand{\DecayF}{{\cal F}}

\newcommand{\MatrixM}{{\cal M}^{\gamma}}

\newcommand{\ETh}{E_{\rm Th}}

\newcommand{\Sqg}{\sqrt{\gamma/D}}
\newcommand{\SqE}{\sqrt{\tilde{\gamma}}}
\newcommand{\Zll}{Z_{1}}
\newcommand{\Zlr}{Z_{2}}
\newcommand{\Zdl}{Z_{4}}

\begin{document}

\DOIsuffix{theDOIsuffix}

\Volume{12}
\Issue{1}
\Month{01}
\Year{2011}

\pagespan{1}{}
\Receiveddate{15 November 2011}
\Accepteddate{2 December 2011}

\title[Transport and Dephasing in a Quantum Dot: \\ Multiply Connected Graph Model]
       {Transport and Dephasing in a Quantum Dot: \\ Multiply Connected Graph Model}

\author[M. Treiber]{Maximilian Treiber\footnote{Corresponding author \quad E-mail: {\sf Maximilian.Treiber@physik.lmu.de}, Phone: +49\,(0)89\,2180\,4533}\inst{1}} 
\address[\inst{1}]{Ludwig Maximilians University, Arnold Sommerfeld Center and Center for Nano-Science, \\ Munich, D-80333, Germany}
\author[O.M. Yevtushenko]{Oleg Yevtushenko\inst{1}}
\author[J. von Delft]{Jan von Delft\inst{1}}

\begin{abstract}
  Using the theory of diffusion in graphs, we propose a model to
  study mesoscopic transport through a diffusive quantum dot.
  The graph consists of three quasi-1D regions: a central region
  describing the dot, and two identical left- and right- wires connected
  to leads, which mimic contacts of a real system. 
  We find the exact solution
  of the diffusion equation for this graph and evaluate the conductance
  including quantum corrections.
  Our model is complementary to the RMT-models describing quantum dots.
  Firstly, it reproduces the universal limit at zero temperature.
  But the main advantage compared to RMT-models is that it allows
  one to take into account interaction-induced dephasing
  at finite temperatures.
  Besides, the crossovers from open to almost closed quantum
  dots and between different regimes of dephasing can be described
  within a single framework. We present results for 
  the temperature dependence of the weak localization correction to 
  the conductance for the experimentally relevant parameter range 
  and discuss the possibility to observe the elusive 0D-regime of 
  dephasing in different mesoscopic systems.
\end{abstract}

\maketitle                   

\section{Introduction}
\label{sect:Introduction}

In the last decades, dephasing in quantum dots has been studied
experimentally and theoretically in great detail. The theoretical
description is largely based on results from random matrix theory (RMT),
emphasizing the universality in the description of a dot, when spatial
degrees of freedom become negligible. While the universal limits are
well understood and reproduced in many experiments, a prediction of the
full temperature dependence of quantities which are sensitive to
dephasing, such as quantum corrections to the classical conductance,
$\Delta g$, are challenging existing theories. Since RMT is not able to
describe the $T$ dependence on its own, several extensions were
introduced in the past to describe their dependence on a dephasing time
$\tPhi$, which has to be included phenomenologically, see
\Sect{sect:DephasingDots} for details.

One of the well-know problems in the theory of dephasing in quantum dots
originated from the predictions of a seminal paper by Sivan, Imry and
Aronov, who showed that dephasing in the so-called 0D regime
($T\ll \ETh$, where $\ETh$ is the Thouless energy), behaves as
$\tPhi \sim T^{-2}$, which results from Pauli blocking of the Fermi sea
\cite{Sivan_QuasiParticleLifetime_1994}. However fundamental the origin
of 0D dephasing is, it has so far not been observed experimentally.
One possible reason for this might be the fact that dephasing is very
weak in this regime, such that quantum corrections may reach their
universal limit $\Delta g \sim 1$.
In general, if the dephasing time is much larger than the time the
electron spends in the dot, $\Delta g$ is governed by a dwelling time
$\, \tDwell \,$ and becomes almost $T$ independent. The remaining small
$T$-dependent part of $\Delta g$ can be masked, for example, by other
$T$-dependent effects coming from contacts or leads. Thus, to facilitate
an experimental observation of 0D dephasing, a comprehensive theory of
transport in the quantum dot connected to leads via some contacts is
needed, which goes beyond the simple picture provided by RMT.

In this paper we propose an alternative to the RMT description
of the quantum dots. Namely, we follow the ideas of
\cite{Doucot_Networks_1985,Doucot_Networks_1986} and model the quantum dot as a network
of 1D wires and use the theory of diffusion in graphs to calculate
$\tPhi$ and $\Delta g$. Earlier papers either focused only on small
graphs, such as 1D rings \cite{Ludwig_Ring_2004,Texier_Ring_2004,Treiber_Dephasing_2009},
or the authors introduced $\tPhi$ only phenomenologically \cite{Doucot_Networks_1985,Doucot_Networks_1986,Texier_NetworksCylinders_2009}.
We generalize the theory of $\tPhi$ for arbitrary graphs and include
the regime $\, T < \ETh $ by taking into account the Pauli principle.
Using this theory, we calculate $\tPhi$ for a network
describing a quantum dot, taking into account effects of the contacts
and the leads. This allows us to demonstrate that the $T^2$-dependence
of the dephasing rate in 0D regime is substantially distorted in usual
transport measurements in quantum dots.

The rest of the paper is organized as follows: In \Sect{sect:DephasingDots}
we give a brief review of known results for
dephasing in quantum dots. In \Sect{sect:DiffusionInGraphs}, basic
results from the theory of diffusion in graphs are presented, and in
\Sect{sect:GraphModel} we will apply this theory to
construct a solvable quantum dot model as an alternative to the
well-known RMT models. Results for the quantum corrections to the
conductance and the dephasing time are presented in the following
Sections. In the conclusions we compare different experimental
setups where 0D dephasing could be observed.

\section{Dephasing in quantum Dots: Brief review of known results}
\label{sect:DephasingDots}

It is well-known that the conductance $g$ of a disordered normal metal
is reduced due to quantum mechanical interference of the electron wave
functions scattered at static impurities. It has been found that the
reduction of $g$ can be expressed via the return probability of coherent
electron paths, $P(\bm{x},\bm{x},t)$, (the so-called {\it Cooperon}) integrated
over time and space \cite{Khmelnitskii_WeakLoc_1984}:
\begin{equation}\label{eq:Deltag}
  \Delta g \equiv g - g_0 = - 4 \ETh
  \int_0^{\infty} \dx{t} \int \dx{^d \bm{x}} \, P(\bm{x},\bm{x},t) \, .
\end{equation}
Here $ \, g_0 \, $ is the classical conductance measured in units of
$\, e^2/h \,$, $ \,  \ETh=D/\Omega^2\, $ is the Thouless energy of the
system, $D$ is the diffusion constant and $\Omega$ is the largest size of the system.
$ \, \Delta g \, $ is usually referred to as the {\it weak localization
correction}.

Quantum coherence is suppressed by a constant magnetic field and by
time-dependent (noisy) fields, or when closed electron paths
contributing to $P(\bm{x},\bm{x},t)$ in \Eq{eq:Deltag} are dephased due to
inelastic scattering events.
The time-scale associated with the latter is called {\it dephasing time}
$\tPhi$. In the absence of other sources of dephasing, $ \, \tPhi \, $
yields an infrared cutoff for the time-integral, \Eq{eq:Deltag}, and
governs the temperature dependence of $\Delta g$ \cite{AAK_Localization_1982}.
At low temperatures, $T \lesssim 1K$, where phonons are frozen, $\tPhi$
is dominated by electron interactions and depends on the dimensionality
$d$ and the geometry of the system. The $T$-dependence of $\tPhi$ in
different regimes is governed by an interplay of $\tPhi$ with the
thermal time $\tTem = 1/T$ and the Thouless time $\tTho = 1/\ETh$,
see \Tab{tab:DephasingTime} for a summary of known regimes in 1D and 2D \cite{Treiber_Noise_2011}.
\begin{vchtable}[htb]
\vchcaption{Dephasing rate $1/\tPhi$ as a function of temperature $T$.}
\label{tab:DephasingTime}\renewcommand{\arraystretch}{1.5}
\begin{tabular}{lccc} \hline
          & $\tTem \ll \tPhi \ll \tTho$ & $\tTem \ll \tTho \ll \tPhi$ & $\tTho \ll \tTem \ll \tPhi$ \\ \hline
1D        & $\propto T^{2/3}$           & $\propto T$                 & $\propto T^{2}$ \\
2D        & $\propto T$                 & $\propto T \Ln{T}$          & $\propto T^{2}$ \\ \hline
\end{tabular}
\end{vchtable}
For low temperatures and small system sizes, when $\ETh$ is the largest
energy scale, dephasing becomes effectively zero-dimensional (0D).
Therefore, it must be relevant for transport in metallic
(diffusive or chaotic) quantum dots \cite{Sivan_QuasiParticleLifetime_1994}.

Note that 0D dephasing requires confinement of the electron paths
during times larger than $\tTho$, since quantum corrections become
$T$ independent for $\tPhi \gg \tTho$ in fully open systems. As an
example, consider the case of a quasi-1D wire of length $L$ connected
to absorbing leads, where $\Delta g$ reads \cite{Akkermans_Book_2007}:
\begin{equation}\label{eq:DeltagWire}
  \Delta g
  = - 4 \sum_{n=1}^{\infty}
    \frac{1}{(\pi n)^2 + \tTho/\tPhi} \Bigl|_{\tPhi \gg \tTho}
  \simeq -\frac{2}{3} \, .
\end{equation}
Thus, a detailed calculation of $\Delta g$ including $\tPhi$ requires
solving the full diffusion equation of the connected quantum dot, which
is hard to achieve analytically for confined systems.

One way to circumvent
this problem is to apply random-matrix theory (RMT) to the scattering
matrix $S$, describing transmission and reflection in the sample.
In such an RMT-model one assumes that the elements of the Hamiltonian
$H$ describing the systems are either real (Gaussian orthogonal ensemble, $\beta_{\rm GOE}=1$) or complex
(Gaussian unitary ensemble, $\beta_{\rm GUE} = 2$) random numbers corresponding to a system with
time-reversal symmetry or broken time-reversal symmetry.\footnote{Note that in this paper, we consider only the spin-less cases.}
Imposing a Gaussian probability distribution $P(H)$, the scattering matrix $S$
can be constructed using so-called R-Matrix theory. Alternatively, a
simpler approach starts from a probability distribution of the
scattering matrix directly, which is of the form $P(S) = {\rm const}$,
and $S$ is again only restricted by symmetry arguments. From the
scattering matrix, the full non-pertubative distribution of the
transmission matrix and the conductance can be obtained. While RMT is
unable to predict the temperature dependence of $\Delta g$ on its own,
the difference in $g$ of the cases $\beta_{\rm GOE}$ and $\beta_{\rm GUE}$ is equivalent
to $\Delta g$ in the limit of $T \to 0$.
Extensions to RMT have been introduced in the past to describe the
dependence of $\Delta g$ on a dephasing time \cite{Brouwer_Dephasing_1997}, e.g. by including a
fictitious voltage probe into the scattering matrix which removes
electrons from the phase-coherent motion of the electrons in the quantum
dot \cite{Buettiker_Coherence_1986}, or by including an imaginary
potential equal to $-i/2\tPhi$ in the Hamiltonian from which the
scattering matrix is derived \cite{McCann_QuantumDots_1996}. It is
expected that $\tPhi$ included in such an approach has the same form as
stated in \Tab{tab:DephasingTime} for $T \ll \ETh$, i.e.
$\tPhi \propto T^{-2}$, but a proof of this expectation and a theory of
a crossover between different regimes is still missing.

\section{Diffusion in graphs}
\label{sect:DiffusionInGraphs}

\begin{figure}[t]
  \centering
  \includegraphics[width=0.9\textwidth]{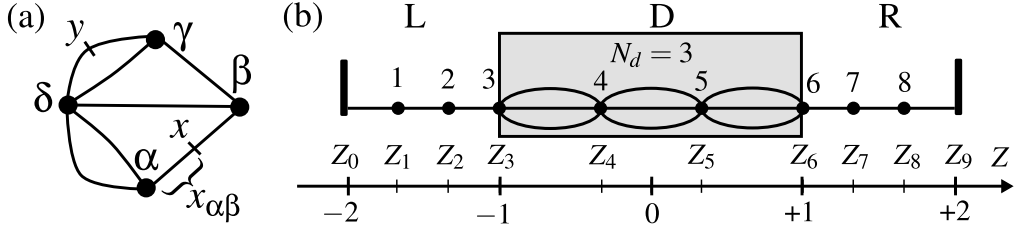}
  \caption{(a) A graph consisting of 9 wires and 6 vertices, denoted by
           Greek letters. (b) A quantum dot realized as a graph with dimensionless
           coordinate $Z=x/L$. The labels $Z_{i}$ denote the
           position of the leads ($i=0,9$) and the vertices
           ($i=1 \dots 8$) on the scale $Z$. Furthermore, the numbers
           $i=1 \dots 8$ correspond to the $i$th row or column of the
           vertex matrix $\MatrixM$, \Eq{eq:MatrixDotModel}}
  \label{fig:Graph}
\end{figure}

In this section, we present basic results from the theory of
diffusion in graphs, following \Ref{Akkermans_Book_2007}.
A graph is defined as a set of quasi-1D wires connected to each other at
vertices, see the example shown in \Fig{fig:Graph}(a).
In this section we will show how the solution to the Laplace transformed
diffusion equation,
\begin{equation}\label{eq:DiffEqLaplace}
  \left( \gamma - D \Delta \right) \Prob_{\gamma}(x,y)
  = \delta(x-y) \, ,
\end{equation}
between arbitrary vertices (with coordinates $ \, x \, $ and $ \, y $)
of such a graph can be obtained. The time-dependent probability, required
to calculate $\Delta g$ and $\tPhi$, can be obtained via an inverse Laplace transform:
\begin{equation}\label{eq:InverseLaplace}
  \Prob(x,y,t) = \frac{1}{2 \pi i}
    \int_{-i\infty}^{+i\infty}\dx{\gamma}
    \, e^{\gamma t} \, \Prob_{\gamma}(x,y) \, .
\end{equation}

It is convenient to introduce the following quantities: We denote the
wire between arbitrary vertices $\alpha$ and $\beta$ as $(\alpha\beta)$ and its length as $L_{\alpha \beta}$.
Furthermore, {\it the running coordinate} along this wire (measured from $\alpha$) 
is denoted $x_{\alpha \beta}$, and in the following, we will not distinguish a vertex from the coordinate
of the vertex on the graph: For example, $P(\alpha,y)$ is equivalent to
$\lim_{x_{\alpha \beta} \to 0} P(x_{\alpha \beta},y)$, for any
neighboring vertex $\beta$ of $\alpha$. The current conservation at some
vertex $ \, \alpha \, $ can be written as follows:
\begin{equation}\label{eq:CurrentConservation}
  - \overline{\sum_{(\alpha \beta)}} \left[ \partial_{x_{\alpha \beta}}
  \Prob_{\gamma}(\mu,x_{\alpha \beta}) \right]_{x_{\alpha \beta} = 0}
  = \delta_{\alpha, \mu} \, ,
\end{equation}
where the symbol $\overline{\sum}_{(\alpha \beta)}$ means summation over
all wires $(\alpha \beta)$ which are connected to $\alpha$.

Consider the point $x$ lying at the coordinate $x_{\alpha \beta}$ of wire
$(\alpha \beta)$ in \Fig{fig:Graph}(a).
The probability to reach $x$ from some
arbitrary other point $y$ of the graph can be expressed in terms of
the probabilities from the neighboring vertices of $x$, i.e.
$\alpha$ and $\beta$:
\begin{equation}\label{eq:LaplaceBond}
  \Prob_{\gamma}(y,x)
  = \frac{\Prob_{\gamma}(y,\alpha)
            \Sh{\Sqg \, (L_{\alpha \beta}-x_{\alpha \beta})}
        + \Prob_{\gamma}(y,\beta)
            \Sh{\Sqg \, x_{\alpha \beta}}
         }{ \Sh{\Sqg \, L_{\alpha \beta}}} \, .
\end{equation}
Validity of the solution (\ref{eq:LaplaceBond}) can be checked directly
by substituting \Eq{eq:LaplaceBond} into \Eq{eq:DiffEqLaplace}.

Inserting \eqref{eq:LaplaceBond} into \eqref{eq:CurrentConservation}
yields the following equations for vertex $\alpha$:
\begin{equation}\label{eq:LinEqs}
  \Prob_{\gamma}(\mu,\alpha) \overline{\sum_{(\alpha\beta)}} \sqrt{\gamma/D} \,
  \Cth{\Sqg L_{\alpha \beta}}
  - \overline{\sum_{(\alpha\beta)}} \Prob_{\gamma}(\mu,\beta)
    \frac{\Sqg}{\Sh{\Sqg L_{\alpha \beta}}}
  = D \delta_{\alpha, \mu}
  \, .
\end{equation}
Writing down \Eq{eq:LinEqs}, for every vertex of the graph, we obtain a set of
linear equations which can be solved for arbitrary vertices. Let us
define a matrix $\MatrixM$ as follows:
\begin{equation}\label{eq:MatrixM1}
  \MatrixM_{\alpha \beta} \equiv \overline{\sum_{(\alpha\delta)}} \left(
  \delta_{\alpha \beta}\Sqg \Cth{\Sqg \, L_{\alpha \delta}}
   - \delta_{\delta \beta} \Sqg \, \Sh{\Sqg L_{\alpha \delta}}^{-1} \right) \, .
\end{equation}
It is easy to check that the diffusion probability between arbitrary
vertices of the graph is given by the entries of the inverse matrix
divided by the diffusion constant
\cite{Texier_NetworksCylinders_2009,Akkermans_Book_2007}:
\begin{equation}\label{eq:ProbM}
  \Prob_{\gamma}(\alpha,\beta)
  = \frac{1}{D} (\MatrixM)^{-1}_{\alpha\beta} \, .
\end{equation}

\section{A graph model for a connected quantum dot}
\label{sect:GraphModel}

In this section we explain how to describe  a connected quantum dot by a
network of 1D wires. The main advantage of this model is that an exact
solution to the diffusion equation can be found.

Consider the network shown in \Fig{fig:Graph}(b). It includes 8 vertices
and describes a quantum dot of total length $2L$ attached via two
contacts of length $L$ to absorbing leads. Multiple wires
connecting the same vertices (e.g. the three wires connecting vertex $4$
with vertex $5$) mimic a larger number of channels.
Below, we use a dimensionless
coordinate $Z=x/L$; the position of the leads is fixed at $Z_0 = -2$,
$Z_9=+2$ and the position of the $3$rd and $6$th vertex, describing the
connection of the dot to the contacts, is fixed at $Z_3 = -1$, $Z_6=+1$. 
The remaining 6 vertices are auxiliary: There are 3 regions in the
system marked by ``L''  (left contact), ``D'' (dot) and  ``R'' (right
contact). We would like to describe diffusion from an arbitrary point in
the system to another. Therefore, we have to place two additional
vertices in each region L, D, R. Positions of these vertices define
running coordinates. They are arbitrary within the corresponding
region, thus each region is subdivided into 3 wires of varying length.
The running coordinates can be expressed via the length of the
connecting wires, e.g. the length of the wire connecting vertices $1$
and $2$ is given by $(\Zlr-\Zll)$.

To describe confinement of the electrons, we assume that all vertices in
the regions L and R (including boundaries) are connected by single wires
while the vertices in the dot (including its boundaries) are connected
via $N_d$ wires. This allows us to tune the system from a simple wire at
$N_d=1$ to an almost closed quantum dot for $N_d\to \infty$. The
corresponding vertex matrix $\MatrixM$, defined in
\Eqs{eq:MatrixM1}, is given by
\begingroup
\renewcommand*{\arraystretch}{1.0}
\begin{equation}\label{eq:MatrixDotModel}
  \MatrixM = 
  \begin{pmatrix}
  \big[\MatrixM_L\big] & \begin{matrix} 0 \\ S_{L} \end{matrix} & \begin{matrix} 0 & 0 \\ \,\,\,\,0\,\,\,\, & \,\,\,\,0\,\,\,\, \end{matrix} & \begin{matrix} 0 \\ 0 \end{matrix} & \begin{matrix} 0 & 0 \\ \,\,0\,\, & 0 \end{matrix} \\
  \begin{matrix} 0 & S_{L} \end{matrix} & C_{LD} & \begin{matrix} S_{LD} & 0 \end{matrix} & 0 & \begin{matrix} \,\,0\,\, & 0 \end{matrix} \\
  \begin{matrix} 0 & \,\,0\,\, \\ 0 & 0 \end{matrix} & \begin{matrix} S_{LD} \\ 0 \end{matrix} & \big[\MatrixM_D\big] & \begin{matrix} 0 \\ S_{DR} \end{matrix} &   \begin{matrix} 0 & 0 \\ \,\,0\,\, & 0 \end{matrix} \\
  \begin{matrix} 0 & \,\,0\,\, \end{matrix} & 0 & \begin{matrix} 0 & S_{DR} \end{matrix} & C_{DR} & \begin{matrix} S_{R} & 0 \end{matrix} \\
  \begin{matrix} 0 & \,\,0\,\, \\ 0 & 0 \end{matrix} & \begin{matrix} 0 \\ 0 \end{matrix} & \begin{matrix} \,\,\,\,0\,\,\,\, & \,\,\,\,0\,\,\,\, \\ 0 & 0 \end{matrix} & \begin{matrix} S_{R} \\ 0 \end{matrix} & \big[\MatrixM_R\big]
  \end{pmatrix} \, .
\end{equation}
\endgroup
We have introduced $ \, 2 \times 2 $ blocks,
\begin{align}
  \MatrixM_L = \MatrixM_{0 1 2 3} \, ,
  \qquad
  \MatrixM_D = N_d \MatrixM_{3 4 5 6} \, ,
  \qquad {\rm and} \qquad
  \MatrixM_R = \MatrixM_{6 7 8 9} \, ,
\end{align}
which are given by:
\begingroup
\renewcommand*{\arraystretch}{1.1}
\begin{align}
  \!\!\!\!\!\!\!\!\!\!\!\!\!
  \MatrixM_{i j k l}= 
  \begin{pmatrix}
  \Cth{\!\SqE[Z_j\!-\!Z_i]} \! + \! \Cth{\!\SqE[Z_k\!-\!Z_j]} & \!\!\! -1/\Sh{\!\SqE[Z_k\!-\!Z_j]} \\
  -1/\Sh{\!\SqE[Z_k\!-\!Z_j]} & \!\!\! \Cth{\!\SqE[Z_k\!-\!Z_j]} \! + \! \, \Cth{\!\SqE[Z_l\!-\!Z_k]} \\
  \end{pmatrix} \, .
\end{align}
\endgroup
Expressions for the entries ``$S$''  and ``$C$'', which correspond to
connected vertices, read
\begin{align}\nonumber
 & S_{L;R}  = -1 / \Sh{\SqE(-1 \mp Z_{\rm 2;7})} \,, \quad  S_{LD;DR} = -N_d / \Sh{\SqE(1 \pm Z_{\rm 4;5}} \,, \\ \nonumber
 & C_{LD;DR} = N_d \Cth{\SqE(1 \pm Z_{\rm 4;5})} + \Cth{\SqE(-1 \mp Z_{\rm 2;7})} \, ,
\end{align}
where we have defined the dimensionless parameter $\tilde{\gamma} = \gamma/\ETh$,
where $\ETh=D/L^2$ is the Thouless energy on the scale $L$.
Note that the total length of the wires which form the graph is
$L_{\rm total} = 2L (N_d + 1)$. Thus, the probabilities obtained via
inversion of the matrix (\ref{eq:MatrixDotModel}), cf. \Eq{eq:ProbM},
are normalized on $L_{\rm total}$. For further calculations, it is more
convenient to change this normalization from $L_{\rm total}$ to the
actual length of the system, $ \, 4 L $: Firstly, we recall that all
$N_d$ wires in the dot connecting the same two vertices have the same
length, i.e. these wires are identical. Consider a point $ {\cal X} $
inside the dot which belongs a given wire (out of $ \, N_d $) and is
infinitesimally close to one of the vertices $\alpha= 4 \ {\rm or} \ 5$. 
The probability to reach $ {\cal X} $ from any other point is equal to
the probability to reach $\alpha$ itself. Let us now introduce a
probability $\ProbD$ to reach $ {\cal X} $ belonging to \emph{any} of
the $ N_d $  wires:
\begin{equation}\label{eq:ProbMDot}
  \ProbD_{\gamma}(\alpha,\beta)
  = N^{(\beta)} \Prob_{\gamma}(\alpha,\beta)
  \equiv N^{(\beta)} \frac{1}{D} (\MatrixM)^{-1}_{\alpha\beta} \, ;
\end{equation}
here $N^{(\beta)} = N_d$ if $\beta$ is a vertex lying in the dot and
$N^{(\beta)} = 1$ otherwise. $\ProbD$ is normalized on $ 4 L $ and it
reflects an enhancement  of the probability for an electron to stay in
the dot by the factor $ \, N_d $.

Furthermore, we define the {\it piecewise continuous} function
$\ProbD_{\gamma}(x,y)$ of {\it continuous} variables $x,y \in [-2L,2L]$
by selecting two appropriate vertices and replacing the wire-length
parameters, $Z_{\alpha}$, by $x/L$ or $y/L$. E.g. the probability to
reach any point $y \in [-L,L]$ in the dot from a point $x \in [-2L,-L]$
in the left contact, is given by $\ProbD_{\gamma}(x,y) = N_d \frac{1}{D} (\MatrixM)^{-1}_{1 4}$
after replacing $\Zll$ by $x/L$ and $\Zdl$ by $y/L$.

\begin{figure}[t]
  \centering
  \includegraphics[width=0.9\textwidth]{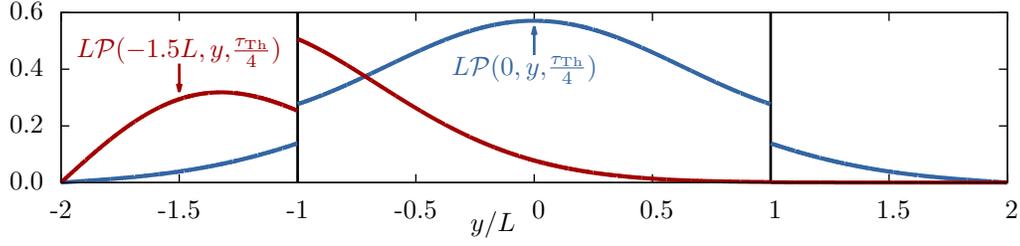}
  \caption{(color online) Probability as a function of space for fixed $t=\tTho/4$,
           $N_d=2$ and initial position $x=-1.5L$ (red curve) or
           $x=0$ (blue curve). The initial positions are marked by
           arrows.}
  \label{fig:Prob}
\end{figure}

An analytic expression for $\ProbD_{\gamma}(x,y)$ can be evaluated
efficiently, but it is lengthy and will be published elsewhere.
Besides, the inverse Laplace transform of $\ProbD_{\gamma}(x,y)$,
cf. \Eq{eq:InverseLaplace}, can be calculated by exploiting the fact
that all poles of $\ProbD_{\gamma}(x,y)$
are simple and coincide with the zeros of the determinant of $\MatrixM$.
Direct calculation yields\footnote{We note
in passing that $S(\tilde{\gamma})$ is proportional to the so-called
spectral determinant, ${\rm det}(-D \Delta + \gamma)$, of the graph
\cite{Akkermans_Book_2007}, implying that it does
not depend on any of the auxiliary coordinates $Z_i$.}
\begin{equation}\label{eq:SDet}
  {\rm det}\MatrixM \propto S(\tilde{\gamma})
  \equiv \Sh{2\SqE}
         \left( (N_d\!-\!1) + (N_d\!+\!1) \Ch{2\SqE} \right) \, .
\end{equation}
Solving equation $ \, S(\tilde{\gamma}) = 0 \, $ yields the following
poles for the graph under consideration:
\begin{equation}\label{eq:SDetZeros}
  \tilde{\gamma}_k = - \left( \frac{k \pi}{2} \right)^2 \, , k \in \mathbb{N}^+ \, ,
  \qquad {\rm or} \qquad
  \tilde{\gamma}_k = -
  \left( k \pi + {\rm arccos}\sqrt{\frac{N_d}{N_d+1}} \right)^2 \, , k \in \mathbb{Z} \, .
\end{equation}
Note that there is no pole at $\tilde{\gamma} = 0$ since the system is open.
Defining the dimensionless function
\begin{equation}
{\cal R}(x,y,\tilde{\gamma}) = \frac{D}{L} \frac{\ProbD_{\tilde{\gamma} \ETh}(x,y) S(\tilde{\gamma})}{S'(\tilde{\gamma})} \, ,
\end{equation}
where $S'(x) = {\partial_x} S(x)$,
we can evaluate the time-dependent probability using the residue theorem
by closing the integral contour in \Eq{eq:InverseLaplace} on the left
half-plane:
\begin{equation}\label{eq:ProbDResidue}
  \ProbD(x,y,t)
  = \frac{1}{L} \sum_{k}
    {\cal R}(x,y,\tilde{\gamma}_k)
    \, {\rm exp} \! \left(\tilde{\gamma}_k \ETh t\right) \, .
\end{equation}
$\ProbD(x,y,t)$ is plotted in \Fig{fig:Prob} for fixed $t=\tTho/4$,
$N_d=2$ and $x$ either in the left contact or in the dot.
We emphasize that for $N_d > 1$, $\ProbD(x,y,t)$ is discontinuous at
$y=\pm L$, describing confinement in the dot. In particular,
$\ProbD(x,y,t) = N_d \ProbD(y,x,t)$ for $x$ in a contact and $y$ in the
dot. Normalization is reflected by the fact that $\ProbD(x,y,t)$
satisfies a semi-group relation
\begin{equation}\label{eq:SemiGroup}
  \int_{-2L}^{2L} \dx{y} \ProbD(x,y,t_1) \ProbD(y,z,t_2)
  = \ProbD(x,z,t_1 + t_2) \, .
\end{equation}
In the next sections we will evaluate the correction to the conductance
and the dephasing time using the probability $\ProbD$.

\section{Quantum corrections to the conductance for the quantum dot model}
\label{sect:QCorrections}

The classical conductance of the system described by
\Eq{eq:MatrixDotModel} is obtained via Kirchhoff's circuit laws,
since the contacts of length $L$ and the central region of length $2L$
(with $N_d$ wires in parallel) are connected in series.
Denoting the contact conductance (i.e. the conductance of the left or
right wire) as $g_c$, we obtain
\begin{equation}\label{eq:GKirchhoff}
  g_0 = \frac{g_c}{2} (1+1/N_d)^{-1} \, .
\end{equation}
Note that the value of $g_c$ cannot be chosen arbitrarily:
Assuming that the substrate, from which the wire (length $L$ and width
$W$) is constructed, is 2D or 3D with mean free path $\ell$ and Fermi
wavelength $\lambda_F$, the conductance is given by
$g_c^{2D} = \ell W / \lambda_F L$ or
$g_c^{3D} = 2\ell W^2 / 3 \pi \lambda_F^2 L$.
Our theory requires $g_c > 4/3$ in order to obtain $g > \Delta g$, and quasi-1D
diffusion requires $\lambda_F \ll \ell,W \ll L$. For a quantum-dot of
the size of several $\mu m$, etched on a GaAs/AlGaAs heterostructure
($\lambda_F \approx 0.05 \mu m$), we can estimate a typical value of
$g_c \sim 5$.

To evaluate the quantum corrections
$\Delta g$, \Eq{eq:Deltag}, we need the return probability defined via
\Eq{eq:ProbMDot} at coinciding $ \, \alpha \, $ and $ \, \beta $. In
this section, we consider the case $ \, T = 0 \, $  (i.e.,
$ \, \tau_{\varphi} \to \infty $) and study $ \, \Delta g \, $ as a 
function of the dissipation parameter $ \, \gamma $. We calculate matrix
elements $[(\MatrixM)^{-1})]_{11}$, $[(\MatrixM)^{-1})]_{44}$ which
yield the return probability for the dot:
\begin{align}\label{eq:RetProbDot}
  \ProbD_{\gamma}(x,x) \Bigl|_{ x \in [-L,L]} = &
  \frac{1}{2 \sqrt{\gamma D} (N_d\!+\!1) S(\tilde{\gamma})} \\\nonumber
& \qquad \times  \left[ (N_d\!-\!1)\Sh{\SqE \frac{x}{L}} - (N_d\!+\!1)\Sh{\SqE\left(\frac{x}{L}\!-\!2\right)} \right] \\\nonumber
& \qquad \times  \left[ (N_d\!+\!1)\Sh{\SqE\left(\frac{x}{L}\!+\!2\right)} - (N_d\!-\!1)\Sh{\SqE \frac{x}{L}} \right] \, ;
\end{align}
and for the left wire:
\begin{align}\label{eq:RetProbLeft}
  \ProbD_{\gamma}(x,x) \Bigl|_{ x \in [-2L,-L]} = &
  \frac{\Sh{\SqE(\frac{x}{L}\!+\!2)}}{2 \sqrt{\gamma D} (N_d\!+\!1) S(\tilde{\gamma})} \\
  & \times \bigg[ (N_d\!-\!1)\bigg( (N_d\!+\!1)\Sh{\SqE \frac{x}{L}}\!+\!(N_d\!-\!1)\Sh{\SqE\left(\frac{x}{L}\!+\!2\right)} \cr
  & \qquad  - (N_d\!+\!1)\Sh{\SqE\left(\frac{x}{L}\!+\!4\right)} \bigg) - (N_d\!+\!1)^2 \Sh{\SqE(\frac{x}{L}\!-\!2)} \bigg] \, ;
\nonumber
\end{align}
respectively. $\ProbD_{\gamma}(x,x)$ for the right wire, $x \in [L,2L]$,
can be obtained from the symmetry property
$\ProbD_{\gamma}(x,x) = \ProbD_{\gamma}(-x,-x)$.
In the limit $\gamma \to 0$, \EqsAnd{eq:RetProbDot}{eq:RetProbLeft} reduce to
\begin{align}\label{eq:RetProbZero}
  \ProbD_{0}(x,x) \Bigl|_{ x \in [-L,L]} \!\!\!= \frac{L((N_d+1)^2 \!-\! (\frac{x}{L})^2)}{2D(N_d+1)} \, , \ 
  \ProbD_{0}(x,x) \Bigl|_{ x \in [-2L,-L]} \!\!\!= \frac{L(2\!+\!\frac{x}{L})(2\!-\!N_d\frac{x}{L})}{2D(N_d+1)} \, .
\end{align}
Note that the return probability diverges for $x \in [-L,L]$ in the limit
$N_d \to \infty$, since the central region is effectively closed in this limit.

\begin{figure}[t]
  \centering
  \includegraphics[width=\textwidth]{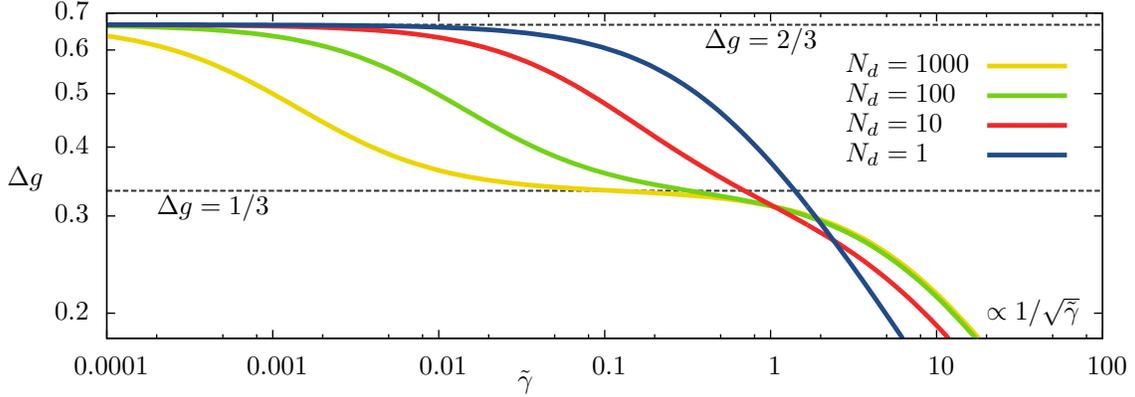}
  \caption{(color online) Dependence of the total quantum correction to the conductance
           of the quantum dot model, Eq.(\ref{eq:Totalg}), on the
           dimensionless dissipation parameter
           $ \, \tilde \gamma = \gamma / E_{{\rm Th}} $.}
  \label{fig:TotalgGamma}
\end{figure}

Similarly to \Eq{eq:GKirchhoff}, the total quantum corrections have to
be properly weighted by using the circuit laws. The total correction can
be written as a sum over all wires $i$ of the network
\cite{Texier_Multiterminal_2004}:
\begin{align}\label{DeltaGtot}
  \Delta g = -4D \frac{1}{\KLen^2}
             \sum_{i} \frac{\partial \KLen}{\partial L_i}
             \int_{\textrm{Wire No.}\ i} {\rm d}x
             \ProbD_{\gamma}(x,x) \, ,
\end{align}
where $\KLen$ is the effective total length of the system obtained
similar to the total resistance. In the case under consideration, we
have
\begin{align}
  \KLen = L_0 + \frac{1}{1/L_1 + \dots + 1/L_{N_d}} + L_{N_d + 1}
        = 2 L (1 + 1/N_d) \, ,
\end{align}
where $L_0 = L$ corresponds to the left wire, $L_{N_d+1} = L$ to the right wire
and $L_1 \dots L_{N_d} = 2L$ to the $N_d$ wires of the dot. We obtain the
following expression for the total quantum correction:
\begin{align}\label{eq:Totalg}
  \Delta g = - \ETh \frac{1}{(1+1/N_d)^2}
             \left[
              \int_{-2L}^{-L} \!\!\!\!\dx{x}\, \ProbD_{\gamma}(x,x) +
              \frac{1}{N_d^2} \int_{-L}^{L} \!\!\!\!\dx{x}\, \ProbD_{\gamma}(x,x) +
              \int_{L}^{2L} \!\!\!\!\dx{x}\, \ProbD_{\gamma}(x,x)
             \right] \, .
\end{align}
In \Fig{fig:TotalgGamma}, we show the total correction to the
conductance according to \Eq{eq:Totalg} as a function of the
dissipation parameter $\gamma$ for different values of $N_d$. Note that
for $\gamma \gg 1$ all curves are $\propto 1/\sqrt{\gamma}$, similar to
an infinite wire with different prefactors corresponding to different

effective wire width. We are mainly interested in the regime
$\gamma \ll 1$, where the main result originates from the left
and right wire and all curves approach the ergodic limit
$ \, \lim_{\gamma \to 0} ( \Delta g ) =2/3 $, cf. \Eq{eq:DeltagWire}.
This limit can be checked in this model by substituting \Eqs{eq:RetProbZero}
into \Eq{eq:Totalg}.
Thus, in the absence of dissipation, our model has qualitatively the
same behavior as RMT theory.
Since the time to reach one contact from the other increases linearly
with $N_d$, there is an intermediate regime at
$ \, 1/N_d < \tilde \gamma < 1 \, $ for large $N_d$, where the system is
described effectively as two wires connected in series via the dot,
which just plays the role of an additional lead, such that
$\Delta g=1/3$.

\section{Evaluation of the dephasing time for the quantum dot model}
\label{sect:Dephasing}

The dephasing time, $ \, \tau_{\varphi} $, can be calculated from the phase difference
acquired by an electron in a time-dependent (fluctuating) 
potential $V(x,t)$ during a time-reversed traversal of its trajectory $x(t)$
\cite{AAK_Localization_1982}:
\begin{equation}\label{PhiV} 
  \Phi[x(\tau)]
  =\int_0^t\dx\tau\,\big[V(x(\tau),\tau)-V(x(\tau),t-\tau)\big]
  \, .
\end{equation}
When averaged over the Gaussian fluctuations of the potential
$\langle e^{i\Phi} \rangle_V = e^{-\frac{1}{2} \langle \Phi^2 \rangle^2}$,
\Eq{PhiV} leads to an exponential cutoff of the return
probability\footnote{Note that in the second equality of
\Eq{eq:ProbDephasing}, we exponentiate the average over closed path, 
see \Ref{Marquardt_Decoherence_2007} for details.}
\begin{equation}\label{eq:ProbDephasing}
  \ProbD(x,x,t)
  \rightarrow \ProbD(x,x,t) \cdot
              \langle e^{i\Phi[x(\tau)]} \rangle_{\{x(\tau)\}}
  \approx     \ProbD(x,x,t) \cdot e^{-\DecayF(x,t)}
\end{equation}
where $ \, \langle \ldots \rangle_{ \{ x(\tau)\} } \, $ means the average is over closed
trajectories $x(\tau)$ of duration $t$, staring and ending at $x$, and
we defined the decay function $\DecayF$ \cite{Marquardt_Decoherence_2007,vonDelft_Decoherence_2007}:
\begin{equation}\label{eq:DecayFInitital}
  \DecayF(x,t) = \int_0^{t} \dx{t_{1,2}} \bigg< \langle VV \rangle(x(t_1),x(t_2),t_1-t_2) - \langle VV \rangle(x(t_1),x(t_2),t-t_1-t_2) \bigg>_{\{x(\tau)\}} \, .
\end{equation}
In the case of the graph model for the quantum dot, the usual operational 
definition of $ \, \tau_{\varphi} \, $ reads
\begin{equation}\label{eq:DephasingDecayF}
   \DecayF(x,\tPhi(x)) = 1 \, ,
\end{equation}
such that the correction to the conductance is given by \Eq{eq:Totalg}
with a position dependent $\gamma(x) = 1/\tPhi(x)$.
The correlation function $\langle VV \rangle$ entering
\Eq{eq:DecayFInitital} is well known for the case of electron
interactions in macroscopically homogeneous disordered systems
\cite{AAK_Localization_1982}. Recently, we have generalized this theory
for inhomogeneous, multiply-connected systems \cite{Treiber_Noise_2011}.
It has been shown that $ \langle VV \rangle $ generically is given by
\begin{equation}\label{eq:CorrFunc}
  \langle VV \rangle(x,y,t) = \frac{4 \pi T}{g_c L} P_0(x,y) \delta_T(t) \, , 
\end{equation}
where $P_0(x,x) = \lim_{\gamma \to 0} P_{\gamma}(x,x)$ and
\begin{equation}
  \delta_T(t) = \pi T w(\pi T t)
                \qquad {\rm with} \qquad w(x)
              = \frac{x\Cth{x}-1}{\ShSq{x}}
\end{equation}
is a broadened $\delta$-function which allows us to take into account
the Pauli principle \cite{Marquardt_Decoherence_2007}.

Inserting \Eq{eq:CorrFunc} into \Eq{eq:DecayFInitital}, we find
\begin{equation}\label{eq:DecayF}
  \DecayF(x,t) = \frac{4 \pi T}{g_c} \int_0^t \! \dx{t_{1,2}} \
          Q(x,t_m,t_M-t_m,t-t_M)
         \left[ \delta_T(t_1-t_2) - \delta_T(t_1+t_2-t) \right] \, ,
\end{equation}
where $t_m = \mint{t_1}{t_2}$ and $t_M = \maxt{t_1}{t_2}$. The function
$Q$ is given by the dimensionless quantity $D P_0/L$, averaged over
closed  random walks:
\begin{equation}\label{eq:DefQ}
  Q(x_0,t_1,t_2,t_3) = \int_{-2L}^{2L} \! \dx{x_{1,2}}
                   \frac{\ProbD(x_0,x_1,t_1) \ProbD(x_1,x_2,t_2) \ProbD(x_2,x_0,t_3)}{\ProbD(x_0,x_0,t_1+t_2+t_3)}
                   \frac{D P_0(x_1,x_2)}{L} \, .
\end{equation}
All probabilities in \Eq{eq:DefQ} can be evaluated analytically from
\Eq{eq:ProbDResidue}, \Eq{eq:ProbMDot} and \Eq{eq:MatrixDotModel}, 
by deriving the corresponding entries in the inverted vertex matrix 
$ \, [ {\cal M}^{\gamma} ]^{-1}$. The integrand is lengthy and we have
chosen the following strategy for calculating the integrals:
1) We use \Eq{eq:ProbDResidue} to rewrite \Eq{eq:DefQ} as:
\begin{eqnarray}\label{eq:QLaplace}
  Q(x_0,t_1,t_2,t_3)
    & = & \sum_{n,k,l} \frac{{\cal Q}(x_0,\gammab_n,\gammab_k,\gammab_l)}
                            {\ProbD(x_0,x_0,t_1+t_2+t_3)}
                       e^{\gammab_n t_1+\gammab_k t_2+\gammab_l t_3}
          \, , \ {\rm with} \\ \label{eq:DefCalQ}
  {\cal Q}(x_0,\gammab_1,\gammab_2,\gammab_3)
    & = & D \int \frac{\dx{x_{1,2}}}{L^3} 
          {\cal R}(x_0,x_1,\gammab_1)
          {\cal R}(x_1,x_2,\gammab_2)
          {\cal R}(x_2,x_0,\gammab_3)
          P_0(x_1,x_2)
          \, .
\end{eqnarray}
The integrals in \Eq{eq:DefCalQ} over space are evaluated symbolically
with the help of a computer algebra program. 2) Since the time
dependence of $Q$ in \Eq{eq:QLaplace} is simply exponential, one of the
time-integrals in \Eq{eq:DecayFInitital} is calculated analytically. 
As a result, $F(x_0,t)$ simplifies to a single time integral and
multiple sums:
\begin{eqnarray}\label{eq:DecayFCompact}
  F(x_0,t) = \frac{(4 \pi)^2}{g_c}
             \frac{
               \displaystyle \sum_{n,k,l}
               {\cal Q}(x_0,\gammab_n,\gammab_k,\gammab_l)
               \int_0^{tT} \!\!\!\! \dx{\tau} \,
               {\cal E}(\tau,\gammab_n,\gammab_k,\gammab_l)
             }{
               \displaystyle \sum_m {\cal R}(x_0,x_0,\gammab_m)
               e^{\gammab_m \ETh t}
             } \, .
\end{eqnarray}
Here, the remaining time dependence of the kernel is incorporated in the
function $ \, {\cal E} $:
\begin{eqnarray}
  & {\cal E}(\tau,\gammab_1,\gammab_2,\gammab_3)
   = w(\pi \tau) e^{c_1 tT}
        \Bigg[
        \frac{\Shs{c_2(tT-\tau)} e^{c_3 \tau }}{2 c_2}
        - \frac{\Shs{\frac{c_3}{2}(tT-\tau)} \Chs{c_2 \tau} e^{ c_3 (tT-\tau)/2}}{c_3}
        \Bigg] , \\
  & {\rm with} \quad c_1 = (\gammab_1 + \gammab_3) \frac{\ETh}{2T}, \ c_2 = (\gammab_1 - \gammab_3) \frac{\ETh}{2T}, \ c_3 = \gammab_2 \frac{\ETh}{T} - c_1 \, .
\end{eqnarray}
3)~The sums and the integral over $ \, \tau \, $ are calculated
numerically.

This strategy allows us to calculate $ \, \tau_{\varphi} \, $ and to
describe the $T$-dependence of $ \, \Delta g \, $ in the quantum 
dot model, including the full crossover between different regimes of
dephasing.

\section{Examples of application}
\label{sect:Application}

\begin{figure}[htb]
  \centering
  \includegraphics[width=0.48\textwidth]{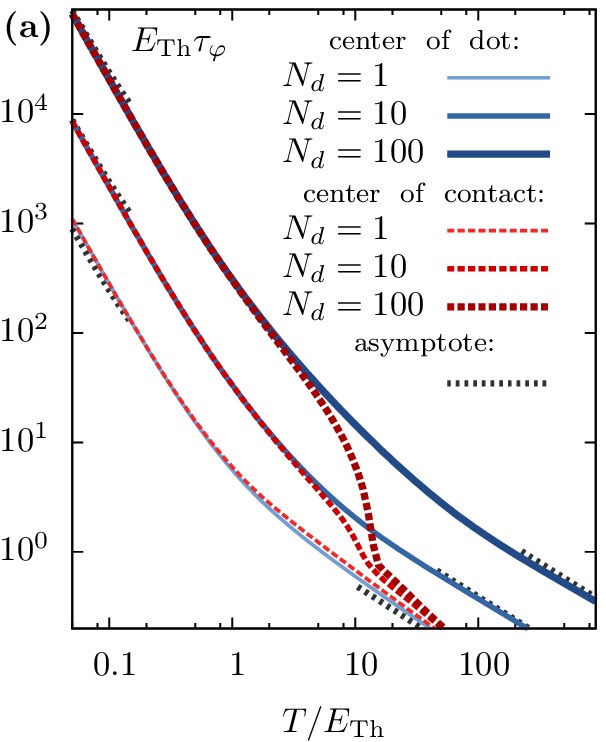}
  \includegraphics[width=0.48\textwidth]{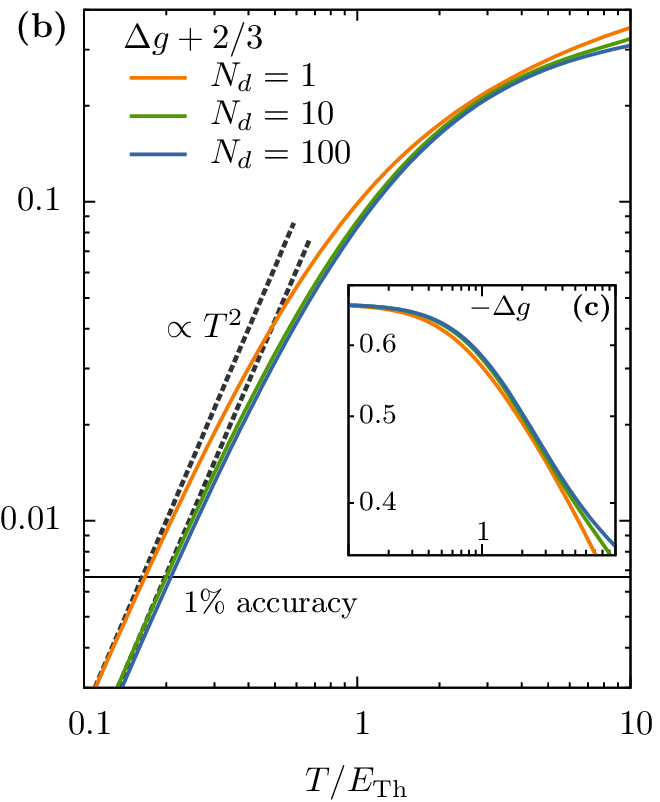}
  \caption{(color online) (a) The dephasing time in units of the Thouless time, $1/\ETh$, plotted
  for several values of $N_d$ and $x_0$. Solid blue lines, correspond
  to $x_0=-0.05L$ close to the center of the dot, while dashed red lines
  correspond to $x_0=-1.55L$ close to the center of the left contact.
  The thickness (and brightness) of the curve determines the number of
  channels in the dot, $N_d=1,10,100$ from thin to thick (and bright to
  dark). The black dotted lines correspond to the asymptotic results,
  \EqsAnd{eq:TPhiFitAAK}{eq:TPhiFit0D},
  derived from an isolated ring geometry, see main text for details.
  (b) The difference $\Delta g + 2/3$ between the correction to the
  conductance, $\Delta g$, and its universal zero-temperature
  value, $\Delta g(T = 0) = -2/3$, plotted as function of temperature.
  0D behavior of the dephasing
  time, characterized by $\Delta g \propto T^2$, appears at very low
  temperatures, requiring a precision much larger than $1\%$ on the
  conductance measurement. Inset: (c) Total correction to the conductance
  $-\Delta g$ (without subtracting  $\Delta g(T = 0)$, plotted as
  function of temperature.}
  \label{fig:tPhi}
\end{figure}

In this section, we use the graph model of the quantum dot to calculate
$ \, \tPhi (x_0,T) \, $ and $ \, \Delta g (T) \, $ in the case
$ \, g_c = 5 \, $ for the parameter $ \, N_d \,  $ ranging from
$ \, N_d = 1 \, $ (no confinement in the central region) to
$ \, N_d = 100 \, $ (almost closed quantum dot connected to ideal leads
via two contacts). Our model is valid for this choice of $ \, g_c $,
see the discussion in \Sect{sect:QCorrections}, and the total
conductance of the system $ \, 1.25 < g_0 < 2.5 \, $ is close to
experimental setups \cite{Huibers_QuantumDots_1998,IleanaRau_DotExperiments_2010}. The results are
shown in \Fig{fig:tPhi}.

{\it The dephasing time} is shown in \Fig{fig:tPhi}(a) for several
values of the origin of the Cooperon, $x_0$, which can belong either to
the central region (solid blue lines) or the contact (dashed red lines). To check the
validity of the results, we compare $ \, \tPhi \, $ at high and small
temperatures with earlier results for an almost isolated quasi-1D ring
of total length $4L$ and total conductance $ \, g_1 $
\cite{Treiber_Book_2010}.

If $ \,\tTem \ll \tPhi \ll \tTho \equiv 1/\ETh$, dephasing is not sensitive to the
boundary conditions and it is described by the theory of infinite
systems \cite{AAK_Localization_1982}. In the ring, the high-$T$ regime
appears at $ \, T \gg g_1 \ETh$.
The formula for $ \, \tau_{\varphi} \, $ in this regime, including sub-leading
terms, reads \cite{Treiber_Book_2010}:
\begin{equation}\label{eq:TPhiFitAAK}
  \frac{\tPhi}{\tTho}
  = \left( \frac{2 g_1 \ETh}{\pi^{3/2} T} \right)^{\frac{2}{3}} \left( 1
    + \frac{2^{\frac{5}{2}}}{3 \pi} |\zeta(1/2)| \left(\frac{\pi^{\frac{3}{2}}}{2 g_1}\right)^{\frac{1}{3}} \left(\frac{\ETh}{T}\right)^{\frac{1}{6}}
    + \frac{2}{9 \sqrt{\pi}} \left(\frac{2 g_1}{\pi^{\frac{3}{2}}}\right)^{\frac{1}{3}} \left(\frac{\ETh}{T}\right)^{\frac{1}{3}}
    \right) \, .
\end{equation}

We have reproduced this high-$T$ behavior in the quantum dot model,
see \Fig{fig:tPhi}(a): Numerically obtained curves
coincide with \Eq{eq:TPhiFitAAK}, after substituting $ \, N_d g_c \, $
for $ \, g_1 $, when $T \gg (g_c N_d) \ETh$.
We note that dephasing in the high-$T$ regime is substantially
inhomogeneous in space, since the relevant trajectories are restricted
to a small region around $x_0$. In particular, for sufficiently high $T$,
all curves for dephasing in the contact ($N_d=1,10,100$) coincide with
the curve for $ \, N_d = 1 $ in the central region, since the number of channels
in the central region is irrelevant for dephasing in the contact. On the other hand,
dephasing in the central region itself becomes weaker with increasing $ \, N_d $, since
$N_d$ increases the effective conductance in this region.

In the low-$T$ regime,\footnote{The intermediate regime, $ \, \tTem \ll \tTho \ll \tPhi  $,
characterized by $\tPhi \propto T^{-1}$ is strongly distorted in the
quantum dot, since: (a) The conductance $g_c$ is relatively small,
reducing the range of validity of this regime, and (b) it occurs when
typical electron trajectories are of the order of the system size making
$\tPhi$ sensitive to the inhomogeneities of the graph.} 
$ \, \tTho \ll \tTem \ll \tPhi $, typical electron trajectories explore
the whole system many times before dephasing becomes
effective \cite{Sivan_QuasiParticleLifetime_1994}. 
The geometry of the system is not important in this case and, therefore,
the low-$T$ regime is usually referred to as {\it the regime of 0D
dephasing}.
In the ring, it occurs at $T \ll \ETh$ with $\tPhi$ given by \cite{Treiber_Book_2010}
\begin{equation}\label{eq:TPhiFit0D}
  \frac{\tPhi}{\tTho}
  = \frac{135 g_1}{32 \pi^2} \left( \frac{\ETh}{T} \right)^{2} \left( 1
    + \frac{16 \pi}{45 g_1} \frac{T}{\ETh}
    + \frac{128 \pi^2}{105} \left( \frac{T}{\ETh} \right)^2
    \right) \, .
\end{equation}
The quantum dot model shows similar behavior at $ \, T \ll \ETh $, 
after
substituting  $g_c N_d$ for the ring conductance. We emphasize that
0D dephasing in our model is governed by atypical trajectories, which
explore the dot and the contacts many times during the time scale
$ \, t \gg \tDwell $. Therefore, the dephasing time is nearly coordinate
independent: Dephasing in the central region and in the contacts is essentially the
same.

{\it The correction to the conductance} is shown in the inset,
\Fig{fig:tPhi}(c), for $N_d=1,10,100$.
We calculated $\Delta g$ from the integral in \Eq{eq:Totalg} with a
position dependent $\gamma(x) = 1/\tPhi(x,T)$. As expected from the
discussion in \Sect{sect:QCorrections}, the curves saturate to the
universal value $\Delta g = -2/3$, when $\tilde{\gamma} \equiv \gamma/\ETh \ll 1/N_d$.
Since $1/\tilde{\gamma} = \tPhi/\tTho \sim (g_c N_d) (\ETh/T)^2$ in this regime and $g_c$ is small
and fixed, saturation occurs when $T \lesssim \ETh$. The intermediate
regime for $1/N_d \ll \tilde{\gamma} \ll 1$, where $\Delta g = -1/3$,
cf. \Fig{fig:TotalgGamma}, is strongly distorted since it lies in the
crossover region between high-$T$ and low-$T$ regime.
We note that at $ \, T < 10 \ETh $, curves for different $ \, N_d \, $
look very similar. Moreover, dephasing is very weak at $ \, T \ll \ETh \, $
where $ \, \Delta g \, $ is governed by a dwell time, $\tDwell$,
of the entire system and is practically $T$-independent. After
subtracting the curve from its universal value, see \Fig{fig:tPhi}(b),
0D dephasing reveals itself as $\Delta g \propto T^2$ for very low
temperatures $T \lesssim 0.2 \ETh$. At $ \, 0.2 \ETh < T < \ETh \, $ one
can observe only a transient, since (i) dephasing is not yet
sufficiently weak to justify $\Delta g \propto T^2$ and (ii) the
0D regime of dephasing is not fully reached, cf. \Fig{fig:tPhi}(a).
Moreover, if the leads are not perfectly absorbing, the transient can be
extended even to lower temperatures due to additional dephasing in
the leads. All this clearly shows that 0D
dephasing cannot be discovered directly in transport measurements
through the quantum dot.
Even at  $T \lesssim 0.2 \ETh$, a fitting of the experimental data would
require $g$ to be measured with a precision of much better than
$1\%$.
Alternative possibilities for the experimental observation
of 0D dephasing are discussed in the Conclusions.

\section{Conclusions}

We have suggested a graph model, which allows one to describe transport
through mesoscopic quantum dots. The graph includes three quasi-1D
regions: identical left- and right- wires and a central region. The
identical wires are connected to ideally absorbing leads and mimic the
contacts of a real system. The number of conducting channels in the
central region can be of the order of- or substantially larger than the
number of channels in the contacts. The latter case corresponds to a
strong confinement of electrons in the central region. Thus the graph
model is able to describe a crossover from opened to closed quantum
dots.

The model which we suggest can be viewed as complementary to the seminal
RMT model. Firstly, the exact solution to the diffusion equation can be
found for the graph model. Secondly, we have shown that our model correctly
reproduces the universal regime of transport in full analogy with
the RMT solution. Even more importantly, the graph model allows us to
take into account interaction induced dephasing in a broad temperature
range, i.e., we can describe the full crossover from 1D to 0D regimes.

Using the solution to the diffusion equation on the graph, we have
described in detail how to calculate the dephasing time and the weak
localization correction to the conductance. Though the intermediate
equations are rather lengthy, we have suggested an efficient combination
of analytical steps (involving computer algebra) and numerical
integration, which helped us to overcome technical difficulties.

The general approach has been illustrated for the system with
$ \, g_c = 5 $. We have demonstrated that 0D dephasing $(\sim T^2)$, which is
governed by the Pauli principle and is very generic, occurs in the
system at $ \, T \ll \ETh \, $ at {\it arbitrary} ratio of the
channel numbers in the dots and leads. In this regime, dephasing is
governed by atypical trajectories which explore the dot and the contacts
many times during the time scale $ \, t \gg \tDwell $ where the
conductance is governed mainly by the dwell time and is almost
$T$-independent. Our results confirm that weak 0D dephasing is
substantially distorted by the influence of the contacts and the leads.
Therefore, its direct experimental observation in transport through the
quantum dot would require not only very low temperatures but also
unrealistically precise measurements. We conclude that
alternative experimental approaches are needed, where either the effects
from the environment are reduced or the system is closed.
One possibility to improve the effective precision of the measurements
is related to extracting $ \, \tPhi \, $ from the $T$-dependence of
the Aronov-Altshuler-Spivak oscillations of the magnetoconductivity in
almost closed mesoscopic rings. This option was discussed in recent
papers \cite{Treiber_Dephasing_2009,Treiber_Book_2010} where all effects
of the environment were taken into account via a constant dwelling
time. We plan to study in more detail the sensitivity of the AAS
oscillations on the distortions from the environment using a ring model
similar to the model of the dot presented here \cite{ToBePublished}.
The other option is to extract $ \, \tPhi \, $ from experimental
measurements of the electric or magnetic susceptibility of \emph{closed}
mesoscopic systems, e.g. by measuring the properties of resonators in
which mesoscopic samples are deposited \cite{Deblock_Polarizability_2002}.
In closed systems, there is no universal limit of the quantum corrections
at $\tPhi \gg \tDwell$, typical for transport through opened systems.
Therefore, the saturation in the closed system can occur at much lower $T$,
making them more suitable for an experimental observation of 0D dephasing.
A theoretical description of such experiments will be published elsewhere.

\begin{acknowledgement}
We acknowledge illuminating discussions with C.~Texier, and support
from the DFG through SFB TR-12 (O.~Ye.), DE 730/8-1 (M.~T.)
and the Cluster of Excellence, Nanosystems Initiative Munich.
\end{acknowledgement}

\end{document}